\documentclass[aps,prl,reprint,showpacs,groupedaddress]{revtex4-1}
\usepackage{amstext}
\usepackage{graphicx}
\usepackage{dcolumn}
\usepackage{bm}

\begin{document}

\title{Demonstration of the Double Penning Trap Technique with a Single Proton}

\author{A. Mooser$^{1,2}$,  S. Braeuninger$^{3,4}$, K. Franke$^{3,5}$, H. Kracke$^{1,2}$, C. Leiteritz$^{1}$,  C. C. Rodegheri$^{3,4}$, H. Nagahama$^{6}$, G. Schneider$^{1,5}$, C. Smorra$^{5}$, K. Blaum$^{3}$, Y. Matsuda$^{6}$, W. Quint$^{4,7}$, J. Walz$^{1,2}$, Y. Yamazaki$^{8}$, and S. Ulmer$^{5}$}
\affiliation{$^1$ Institut f\"ur Physik, Johannes Gutenberg-Universit\"at D-55099 Mainz, Germany}
\affiliation{$^2$ Helmholtz-Institut Mainz,  D-55099 Mainz, Germany}
\affiliation{$^3$ Max-Planck-Institut f\"ur Kernphysik, Saupfercheckweg 1, D-69117 Heidelberg, Germany}
\affiliation{$^4$ Ruprecht Karls-Universit\"at Heidelberg, D-69047 Heidelberg, Germany}
\affiliation{$^5$ RIKEN, Ulmer Initiative Research Unit,  2-1 Hirosawa, Wako, Saitama 351-0198, Japan}
\affiliation{$^6$ Graduate School of Arts and Sciences, University of Tokyo, Tokyo 153-8902, Japan}
\affiliation{$^7$ GSI - Helmholtzzentrum f\"ur Schwerionenforschung, D-64291 Darmstadt, Germany}
\affiliation{$^8$ RIKEN, Atomic Physics Laboratory,  2-1 Hirosawa, Wako, Saitama, 351-0198, Japan}

\author{{submitted:} {2013/04/02} -- {accepted for publication:} {2013/05/05}}

\begin{abstract}
Spin flips of a single proton were driven in a Penning trap with a homogeneous magnetic field. For the spin-state analysis the proton was transported into a second Penning trap with a superimposed magnetic bottle, and the continuous Stern-Gerlach effect was applied. This first demonstration of the double Penning trap technique with a single proton suggests that the antiproton magnetic moment measurement can potentially be improved by three orders of magnitude or more.
\end{abstract}

\maketitle

\begin{figure}[htb]
        \centerline{\includegraphics[width=9cm,keepaspectratio]{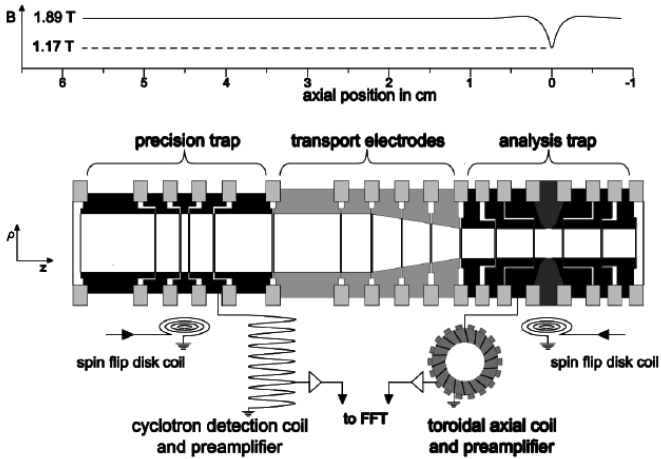}}
            \caption[FEP]{Schematic of the double Penning trap setup. The system consists of two Penning traps which are connected by transport electrodes. The central ring electrode of the analysis trap is made of ferromagnetic material. For further details see text.}
            \label{fig:TrapDetect}
\end{figure}

\begin{figure}[htb]
        \centerline{\includegraphics[width=9cm,keepaspectratio]{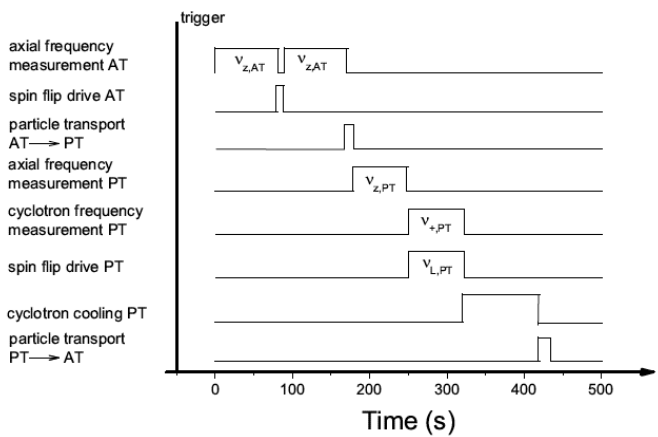}}
            \caption[FEP]{Illustration of a double Penning trap sequence. For details see text.}
            \label{fig:DoubleTrap}
\end{figure}

CPT symmetry is a fundamental symmetry in the local, relativistic quantum field theories involved in the Standard Model \cite{CPT}. It implies exact equality between the fundamental properties of particles and their corresponding antiparticles. Any measured difference between masses, charges, lifetimes or magnetic moments of an exactly conjugated matter/antimatter system would thus indicate physics beyond the Standard Model, and potentially contribute to the solution of one of the hottest topics in modern physics -- the observed matter/antimatter asymmetry in the universe \cite{Dine}. This is the fascination and motivation driving experiments which compare the properties of matter and antimatter at lowest energies  and greatest precision \cite{VanDyck,JerryAntiproton,Enomoto,HORI,ALPHASpinFlips,ATRAPTrap,AEgIS}.\\
Our experiments \cite{UlmerPRL,CCR, MooserPRL} focus on the precise comparison of the magnetic moments of the proton p and the antiproton $\bar{\text{p}}$, $\mu=g/2\cdot q \hbar / (2 m)$, where $g$ is the so-called \emph{g}-factor, and $q / m$ the charge-to-mass ratio. The magnetic moment in units of the nuclear magneton $\mu_{N}$

    \begin{eqnarray}
    \frac{\mu}{\mu_N}=\frac{g}{2}=\frac{\nu_L}{\nu_c}~,
    \label{eq:MagMom}
    \end{eqnarray}
can be determined by measuring the free cyclotron frequency $2\pi\nu_\text{c}= q/m\cdot B_0$ and the spin precession frequency $2\pi\nu_L=g/2\cdot q/m\cdot B_0$ (also called Larmor frequency) of a single particle stored in a cryogenic Penning trap, where $B_0$ is the trap's magnetic field. The cyclotron frequency is determined by image current detection, whereas for the measurement of $\nu_L$ the continuous Stern-Gerlach effect is applied \cite{DehmeltCSG}.
\begin{figure}[htb]
        \centerline{\includegraphics[width=7cm,keepaspectratio]{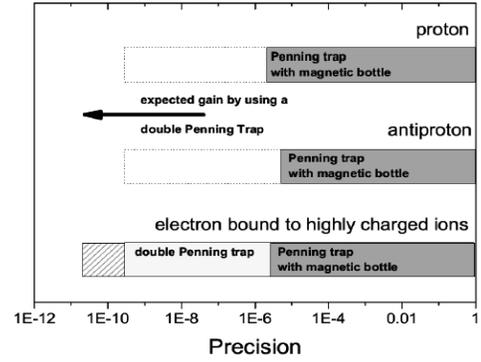}}
            \caption[FEP]{Precision achieved in recent measurements of magnetic moments with single trapped particles.  The dark grey bars indicate the precision which has been achieved in Penning traps with a superimposed magnetic bottle for spin detection.  Light grey bar: Using the double Penning trap technique the precision improved dramatically to about 0.5\,ppb in experiments with  a single electron bound to a highly charged ion. Hatched bar: a phase sensitive detection method recently enabled a 20\,ppt measurement \cite{SvenPNA}. A gain in precision by more than three orders of magnitude is expected from using the double Penning trap technique. }
            \label{fig:Compare}
\end{figure}
In this scheme a strong magnetic bottle $B(z)=B_0+B_2z^2$, where $z$ is the direction of the magnetic field, is superimposed on the Penning trap. This couples the spin-magnetic moment of the particle to its axial oscillation frequency $\nu_z$, and thus, reduces the determination of the spin direction to a non-destructive frequency measurement. A magnetic radio frequency field at $\nu_\text{rf}$ is applied to induce spin transitions \cite{Rabi1}, and from a measurement of the spin flip probability $P_\text{SF}(\nu_\text{rf})$, the Larmor frequency $\nu_L$ is obtained. By using this method the magnetic moment of a single trapped proton was measured by two independent groups with precisions of 8.9$\cdot10^{-6}$ \cite{CCR} and 2.5$\cdot10^{-6}$ \cite{Jack1}.  Applying the same techniques  to the antiproton, diSciacca \emph{et al.} recently reported on a 4.4~ppm measurement of the antiparticle's magnetic moment \cite{Jack2}. Compared to the former value \cite{Pask1}, this improves the precision of $\mu_{\bar{\rm{p}}}$ by an unprecedented factor of 680.
However, the achieved experimental precision is limited by the strong magnetic bottle $B_2\approx 300\,000\,$T/m$^2$ which broadens the Larmor resonance line significantly \cite{Brown2}. To overcome this problem, a GSI/University of Mainz collaboration developed the elegant double Penning trap method \cite{häffner2003double}, where the analysis of the spin state and the precision measurements of $\nu_c$ and $\nu_L$ are separated to two traps: an analysis trap (AT) with the superimposed $B_2$, and a precision trap (PT), in which the magnetic field is a factor of $75\,000$ more homogeneous. This reduces the line width of the spin resonance by orders of magnitude and boosts experimental precision. Figure~\ref{fig:Compare} shows that this method has been applied with great success in measurements of the magnetic moment of the electron bound to highly charged ions, where sub-ppb precision was achieved \cite{Hermanspahn,Hartmut,Verdu,Sven}. So far, in case of the (anti)proton magnetic moment measurements the application of this method was not possible. \\
In this Letter we report the first successful demonstration of the double Penning trap method with a single trapped proton. This is a major step towards a dramatic improvement in the measurement precision of the magnetic moment of both the proton and the antiproton to the ppb level and beyond. \\
The double Penning trap setup is shown in Fig.~\ref{fig:TrapDetect} \cite{CCR}. This assembly is mounted into the horizontal bore of a temperature and pressure stabilized superconducting magnet, with a field strength of $B_0=$1.89$\,$T. The precision trap is located in the homogeneous center with $\Delta B / B \approx 10^{-6}$ in a volume of about 1$\,$cm$^3$.
The two traps consist of carefully designed stacked cylindrical electrodes, which are made out of oxygen free copper. All electrodes are gold-plated, which prevents oxidation of the inner trap surfaces. The PT has an inner diameter of 7~mm, whereas the one of the AT is only 3.6~mm. The central ring electrode of this small trap is made of ferromagnetic CoFe material, which has a saturation magnetization of $B_\text{sat}=2.25\,$T. Together with the small diameter of the AT this material produces the strong magnetic bottle. The distance between the trap centers is 45$\,$mm. The residual magnetic inhomogeneity $B_2$ in the center of the precision trap is $4\,$T/m$^2$. The traps are connected by transport electrodes, which are used to shuttle the single particle adiabatically back and forth between AT and PT. Small coils are mounted close to the Penning trap electrodes. Strong rf drives applied to these coils at frequencies close to the Larmor frequency $\nu_L\approx 81\,$MHz induce spin transitions. Although the copper electrodes act as efficient rf shields, the field amplitude $b_\text{rf}$ which penetrates through the 140~$\mu$m slits between adjacent electrodes is still strong enough to achieve Rabi frequencies $\Omega=\mu_p b_\text{rf}/\hbar$ of several 10$\,$Hz. The whole assembly shown in Fig.$\,$\ref{fig:TrapDetect} is mounted in a hermetically sealed, pinched-off, and cryo-pumped vacuum chamber with a total volume of about one liter. In this trap-can pressures in the order of 10$^{-16}\,$mbar are achieved which avoids particle loss \cite{WolfgangGerry}. \\
A single proton in the precision trap oscillates perpendicular to the magnetic field at the circular modified cyclotron frequency $\nu_+$ of about 28.9$\,$MHz. The electrostatic potential slightly shifts $\nu_+$ from $\nu_c$ \cite{Brown}. In addition to this radial motion an axial oscillation along the magnetic field lines occurs at $\nu_z=623\,$kHz, which is due to the axial electrostatic potential of the Penning trap. A magnetron motion at $\nu_-=6.7\,$kHz is induced by the crossed magnetic and electric fields. It is also perpendicular to $B_0$.  \\
To measure $\nu_+$ and $\nu_z$ of the single trapped proton, highly sensitive detection systems are used. Small image currents $I_p$ are induced in the trap electrodes by the particle's motion and are picked up by detection coils with inductances $L$ and high quality factors $Q$. Together with the trap capacitance $C_T$ they form a tuned circuit. On resonance $\nu_r$ each detection system acts as an effective parallel resistance $R_p = 2\pi \nu_r L Q $. When the particle's respective oscillation frequency and $\nu_r$  are matched, a voltage drop $u_p=R_pI_p$ is obtained, which is amplified and analyzed with a fast Fourier transform (FFT) spectrum analyzer. A particle excited to high oscillation amplitudes appears as a peak in the FFT spectrum. The interaction with the detection resistor $R_p$ cools the proton resistively with a cooling time constant $\tau=m/R_p\cdot D^2/q^2$, where $D$ is a characteristic trap dimension. When cooled to thermal equilibrium, the particle acts like a series tuned circuit, which shorts the thermal noise  \cite{JohnsonNoise,NyquistNoise} of the detection system, and appears as a dip in the FFT spectrum \cite{Wine}. For a non-destructive frequency measurement such a particle dip is recorded and the frequencies $\nu_z$ and $\nu_+$ are obtained via a best fit to the data. In contrast to this direct detection method the magnetron frequency is measured via sideband coupling \cite{Cornell}. This method can also be used to measure $\nu_+$ \cite{UlmerPRL2}. \\
For an application of the double Penning trap technique the sequence shown in Fig.~\ref{fig:DoubleTrap} is applied. First, a single proton is prepared in the AT and its spin state is determined. To this end, the axial frequency is measured, and afterwards a spin flip drive is applied with an amplitude sufficiently large to saturate the incoherent spin-transition to $P_\text{SF}=50\,\%$ \cite{UlmerPRL}. Subsequently the axial frequency is measured again. A spin quantum jump in our magnetic bottle shifts the axial frequency by 171$\,$mHz. Once such a frequency jump is identified  \cite{MooserPRL}, the spin state is defined.
Next, the single proton is transported to the PT, and while a spin-flip drive is applied, the cyclotron frequency is measured. The experimental cycle is finished by transporting the particle back to the AT where the spin state is analyzed as above. In this way a signature is obtained for whether or not the drive applied in the PT flipped the spin.\\
Obviously, for the application of this technique, ``single spin flip resolution'' is required \cite{MooserPRL}, which means, that the spin state of the proton has to be identified with high fidelity $F$, where $F$ is the fraction of correctly identified spin states. This is very difficult because of the strong magnetic bottle $B_2$, which also couples the magnetic moment of the orbital angular magnetic momentum to the axial oscillation frequency. Small fluctuations of the energies in the radial modes, $E_+$ and $E_-$, cause axial frequency fluctuations, which can be expressed as
    \begin{eqnarray}
    \Delta\nu_z(n_+,n_-)&=&\nonumber\\
    \frac{h \nu_+}{4\pi^2 m_\text{p}\nu_z}\frac{B_2}{B_0}&\cdot&\left(\left(n_++\frac{1}{2}\right)+\frac{\nu_-}{\nu_+}\left(n_-+\frac{1}{2}\right)\right)~,
    \label{eq:Bottleshift}
    \end{eqnarray}
where $n_+$ is the cyclotron and $n_-$ the magnetron quantum number, respectively.
A cyclotron quantum jump $\Delta n_+=\pm 1$ causes an axial frequency shift of $\Delta\nu_z=\pm63\,$mHz and a transition of the magnetron quantum number $\Delta n_-=\pm 1$ leads to $\Delta\nu_z=\pm54\,\mu$Hz. Thus, to clearly observe spin transitions it is crucial to keep axial frequency fluctuations induced by radial quantum transitions small in comparison to the frequency shift $\Delta\nu_\text{z,SF}=171\,$mHz induced by a spin flip. \\
The transition amplitude for electric dipole transitions between cyclotron quantum states is
    \begin{eqnarray}
    \Gamma_{i\rightarrow f}=qE_0\sqrt{\frac{\hbar}{2\pi m_p\nu_+}\frac{n_+}{2}},
    \label{eq:Transition}
    \end{eqnarray}
where $E_0$ is the electrical field amplitude of a spurious noise-drive. The transition rate $\delta n_+/dt$ scales as $\Gamma_{i\rightarrow f}^2$, and is proportional to the cyclotron quantum number $n_+$. Thus, to resolve single proton spin flips it is of major importance to cool the particle to low cyclotron energies $E_+$. \\
The scaling of the axial frequency $\nu_z(E_+)$ in the strong magnetic bottle (see eq.~\ref{eq:Bottleshift}) enables a calibration of the absolute modified cyclotron energy $E_+$. For such an energy calibration the particle is transported to the precision trap where it interacts with the cyclotron detector \cite{ulmer2011Det}. With a typical noise correlation-time of $\tau_+=40\,$s the particle's cyclotron mode is thermalized, which means that $E_+$ follows a random walk in thermal energy space. The distribution of the energy values $E_+$ taken on in this process is described by the probability function $w(E_+)=\frac{1}{k T_+}\exp(-E_+/(k_B T_+))$. After thermalization the particle is shuttled to the analysis trap, and the axial frequency $\nu_z$ is measured. Then, the single proton is transported back to the PT and thermalized again. This cycle is repeated for hundreds of times.
\begin{figure}[htb]
        \centerline{\includegraphics[width=8cm,keepaspectratio]{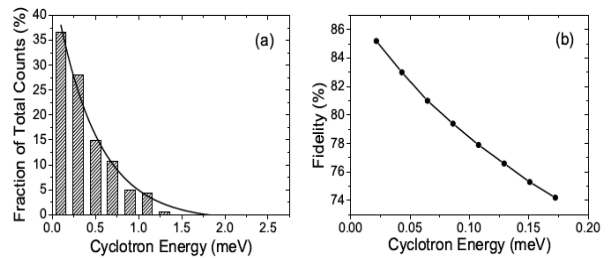}}
            \caption[FEP]{a.) Histogram of axial frequency measurements after thermalization of the modified cyclotron mode with the detection system it the PT. The abscissa is scaled to absolute energy values. For further details see text. b.) Single spin flip fidelity as a function of the cyclotron energy $E_+$. Note the change in the cyclotron energy scale. }
            \label{fig:Transition}
\end{figure}
Figure \ref{fig:Transition} a.) shows the results obtained from such a sequence plotted as a histogram. Together with a fit of $w(E_+)$ to the data and an independent measurement of $B_2$ \cite{CCR}, the measured frequencies are scaled to absolute energy-values (abscissa of histogram). These results constitute a measurement of the temperature of the cyclotron detector $T_+$, probed with the single particle. For this data-set $T_+ = 5.3\,$K is obtained. After such a detailed temperature calibration the cyclotron energy $E_+$ of the single particle can be determined via a single axial frequency measurement. \\
By using this temperature calibration the axial frequency fluctuation $\Xi$, defined as the rms value of the differences of two subsequent axial frequency measurements $\nu_{z,i}- \nu_{z,i+1}$ \cite{UlmerPRL}, was recorded for different $E_+$-values. Based on this result the spin flip fidelity $F$, is obtained as a function of $E_+$ using simulated axial frequency noise data. Results are shown in Fig.~\ref{fig:Transition} b.). At cyclotron energies below 150$\,\mu$eV a spin-flip fidelity above 75$\,\%$ is obtained, which means, that three out of four induced spin flips can be clearly identified. With the current cyclotron detector, the average probability to prepare a particle which meets these requirements is about $20\,\%$, and the preparation of such a particletakes on average about 2$\,$h.\\
For the application of the double Penning trap technique a single proton is prepared at low $n_+$  in the AT. Significant Larmor frequency shifts arise from large magnetron radii. Thus, in the next step the latter is cooled to about 10$\,\mu$m by coupling to the axial mode \cite{Cornell}. This enables reliable saturation of the spin transition to 50$\,\%$ spin flip probability, which is necessary for a high-fidelity spin state detection \cite{MooserPRL}. When the spin-state is defined, the particle is transported to the precision trap, where the free cyclotron frequency is obtained by determining  $\nu_+$ and $\nu_z$. The magnetron frequency  $\nu_-$  is measured routinely only once a day. In parallel to the measurement of the modified cyclotron frequency $\nu_+$, spin flips are driven in the same way as in the AT. The measurement of $\nu_+$ heats the cyclotron mode, and thus, in a next step the proton is thermalized with the cyclotron detector, subsequently transported back to the AT, and the cyclotron energy is determined. In case $E_+ > 150\,\mu$eV, which corresponds to a single spin flip fidelity $F<75\,\%$, the particle is shuttled back to the PT and the cyclotron mode is thermalized again. Once $E_+$ is low enough to resolve single spin flips with fidelity $F>75\,\%$, the spin eigenstate is determined. From this measurement we deduce whether the spin in the PT has flipped. For a determination of the spin-flip probability in the PT this sequence is repeated many times. Data were taken for a resonant and an off-resonant spin-flip drive in the PT, the latter detuned by 1$\,$MHz. Figure~\ref{fig:FFF} shows the
\begin{figure}[htb]
        \centerline{\includegraphics[width=9cm,keepaspectratio]{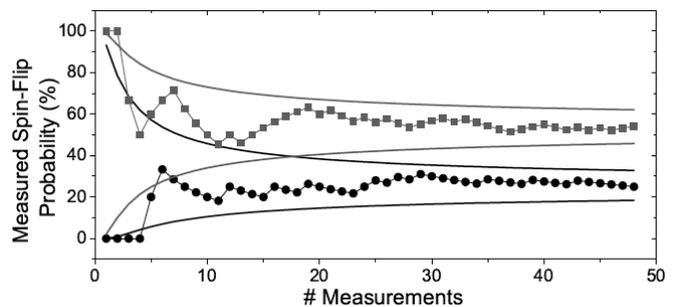}}
            \caption[FEP]{Left: Spin flip probability in the PT as a function of measurement time. The squares represent data points where a resonant spin flip drive was applied, while the dots represent results of a reference measurement, where the drive was detuned by 1$\,$MHz. The solid lines indicate the 1$\sigma$ confidence bands.}
            \label{fig:FFF}
\end{figure}
evolution of the cumulative spin flip probability for the resonant as well as the off-resonant drive as a function of the number of measurements. The solid lines indicate the 1$\sigma$ confidence intervals. Both measurements converge to constant values. The off resonant data point reaches $P_\text{SF} \approx 25\,\%$ since one out of four spin flip trials in the precision trap is misinterpreted. This is caused by the limited fidelity $F$ for the spin state detection, which results in a measured spin flip probability $P_\text{SF,m}=P_\text{SF,t}\cdot F+(1-P_\text{SF,t})\cdot(1-F)$, where $P_\text{SF,t}$ is the actual spin flip probability. For the resonant data point $P_\text{SF} \approx 52(8)\,\%$ are measured. The significant shift of 27$\,\%$ between the two measurements proves the unambiguous detection of proton spin quantum jumps driven in the homogeneous magnetic field of the PT. A scan of the full Larmor resonance would constitute a first proton \emph{g}-factor measurement by means of the double Penning trap technique. So far, this method has only been applied to measure electron magnetic moments, however, the proton magnetic moment is about a factor of 650 times smaller and application of this method is much more challenging. \\
In conclusion, for the first time the application of the double Penning trap method has been demonstrated with a single proton. This result paves the way towards the first direct high precision measurement of the proton and antiproton magnetic moment in the BASE experiment at the Antiproton Decelerator of CERN \cite{AD1,AD2}. \\

We acknowledge financial support of RIKEN Initiative Research Unit Program, the Max-Planck Society, the BMBF, the Helmholtz-Gemeinschaft, HGS-HIRE, and the EU (ERC Grant
No. 290870-MEFUCO).

\end{document}